\documentclass[twocolumn]{svjour3}         
\smartqed  
\usepackage[compress]{cite}
\usepackage{graphicx}
\begin{document}
\newcommand{\REV}[2]{{#2}}
\title{Lattice Boltzmann simulations in microfluidics: probing the no-slip
boundary condition in hydrophobic, rough, and surface nanobubble laden microchannels}


\author{Jens Harting\and Christian Kunert\and Jari Hyv\"aluoma
}

\date{Received: date / Accepted: date}

\institute{
Jens Harting\at
Department of Applied Physics, TU Eindhoven\\
Den Dolech 2, 5600MB Eindhoven, The Netherlands\\
Institute for Computational Physics, University of Stuttgart,
Pfaffenwaldring 27, 70569 Stuttgart, Germany\\
\email{j.harting@tue.nl}
\and
Christian Kunert\at
Institute for Computational Physics, University of Stuttgart,
Pfaffenwaldring 27, 70569 Stuttgart, Germany\\
\email{kuni@icp.uni-stuttgart.de}
\and
Jari Hyv\"aluoma\at
Department of Physics, University of Jyv\"askyl\"a, FI-40014
Jyv\"askyl\"a, Finland\\
\email{jari.hyvaluoma@jyu.fi}
}

\maketitle


\begin{abstract}
\REV{ED}{In this contribution we review recent efforts on investigations of the effect
of (apparent) boundary slip by utilizing lattice Boltzmann simulations. We demonstrate the applicability of the
method to treat fundamental questions in microfluidics by investigating fluid
flow in hydrophobic and rough microchannels as well as over surfaces covered by nano- or microscale gas bubbles.
}
\end{abstract}

\section{Introduction}
\REV{ED}{During the last few decades the miniaturization of technical devices down to
submicrometric sizes has made considerable progress. In particular, so-called
micro\-electro-mechi\-anical systems (MEMS) became available for chemical,
biological and technical applications leading to the rise of ``microfluidics''
about 20 years ago~\cite{bib:tabeling-book}. A wide variety of microfluidic
systems including gas chromatography systems, electrophoretic separation
systems, micromixers, DNA amplifiers, and chemical reactors were developed.
Next to those ``practical applications'', microfluidics was used to answer
fundamental questions in physics including the behavior of single molecules or
particles in fluid flow or the validity of the no-slip boundary
condition~\cite{bib:tabeling-book,bib:lauga-brenner-stone-05}.  The latter is
the focus of the current review and is investigated in detail by mesoscopic
computer simulations.}

Reynolds numbers in microfluidic systems are usually small, i.e., usually below
0.1. In addition, due to the small scales of the channels, the surface to
volume ratio is high causing surface effects like wettability or surface
charges to be more important than in macroscopic systems. Also, the mean free
path of a fluid molecule might be of the same order as the characteristic
length scale of the system. For gas flows, this effect can be characterized by
the so-called Knudsen number~\cite{knudsen09a}. While the Knudsen number
provides a good estimate for when to expect rarefaction effects in gas flows,
for liquids one would naively assume that its velocity close to a surface
always corresponds to the actual velocity of the surface itself. This
assumption is called the no-slip boundary condition and can be counted as one
of the generally accepted fundamental concepts of fluid mechanics. However,
this concept was not always well accepted. Some centuries ago there were long
debates about the velocity of a Newtonian liquid close to a surface and the
acceptance of the no-slip boundary condition was mostly due to the fact that no
experimental violations could be found, i.e., a so-called boundary slip could
not be detected. 

In recent years, it became possible to perform very well controlled experiments
that have shown a violation of the no-slip boundary condition in sub-micron
sized geometries. Since then, mostly
\REV{3}{experimental~\cite{bib:lauga-brenner-stone-05,craig-neto-01,bib:tretheway-meinhart-04,cheng-giordano-02,choi-westin-breuer-03,baudry-charlaix-01,bib:cottin-bizone-etall-02,vinogradova-yakubov-03,bib:tretheway-meinhart-04},
but also theoretical works~\cite{vinogradova-95,bib:degennes-02}, as well as
computer
simulations~\cite{succi02,bib:barrat-bocquet-99,bib:cieplak-koplik-banavar-01,bib:thompson-troian-1997,bib:tretheway-zhu-petzold-meinhart-2002}
have been performed to improve our understanding of boundary slip.} The topic is
of fundamental interest because it has practical consequences in the physical
and engineering sciences as well as for medical and industrial applications.
Interestingly, also for gas flows, often a slip length much larger than
expected from classical theory can be observed.  Extensive reviews of the slip
phenomenon have recently been published by Lauga et
al.~\cite{bib:lauga-brenner-stone-05},  Neto et al.~\cite{neto-etal-05}\REV{ED}{, as well as Bocquet and Barrat~\cite{bib:bocquet-barrat-2007}.}

The reason for such findings is that the behavior of a fluid close to a solid
interface is very complex and involves the interplay of many physical and
chemical properties. These include the wettability of the solid, the shear rate
or flow velocity, the bulk pressure, the surface charge, the surface roughness,
as well as impurities and dissolved gas.  Since all those quantities have to be
determined very precisely, it is not surprising that our understanding of the
phenomenon is still very unsatisfactory.
Due to the large number of different parameters, a significant dispersion
of the results can be observed for almost similar
systems~\cite{bib:lauga-brenner-stone-05,neto-etal-05}. For example,
observed slip lengths vary between a few
nanometres~\cite{bib:churaev-sobolev-somov-84} and
micrometers~\cite{bib:tretheway-meinhart-04} and while some authors find a
dependence of the slip on the flow
velocity~\cite{choi-westin-breuer-03,bib:zhu-granick-01,craig-neto-01}, others
do not~\cite{cheng-giordano-02,bib:tretheway-meinhart-04}.

A boundary slip is typically quantified by the so-called slip length
$\beta$ -- a concept that was already proposed by Navier in 1823. He
introduced a boundary condition where the fluid velocity at a surface is
proportional to the shear rate at the surface~\cite{bib:Navier} (at
$x=x_0$), i.e., 
\begin{equation}
v_z(x_0)=\beta\frac{\partial v_z(x)}{\partial x}. 
\end{equation}
In other words, the slip length $\beta$ can be defined as the distance
from the surface where the relative flow velocity vanishes.
Assuming a typical Poiseuille setup consisting of a pressure
driven flow of an incompressible liquid between two infinite planes, the
velocity in flow direction ($v_z$) at position $x$ between the planes is
given by
\begin{equation} 
\label{eq:plattenprofil} 
v_z(x)=\frac{1}{2 \mu}\frac{\partial P}{\partial z}
\left[ d^2-x^2-2d\beta  \right],
\end{equation}
where $2d$ is the distance between the planes, and $\mu$ the dynamic viscosity.
${\partial P}/{\partial z}$ is the pressure gradient. In
contrast to a no-slip formulation, the last term in
Eq.~\ref{eq:plattenprofil} linearly depends on the slip length $\beta$.

Most recent computer simulations apply molecular dynamics and report
increasing slip with decreasing liquid
density~\REV{ED}{\cite{bib:koplik-banavar-willemsen-89,bib:thompson-robbins-1990}}
or liquid-solid
interactions~\REV{ED}{\cite{bib:cieplak-koplik-banavar-01,bib:nagayama-cheng-2004}},
while slip decreases with increasing
pressure~\cite{bib:barrat-bocquet-99}. These simulations are usually
limited to a few tens of thousands of particles, length scales of a few
nanometres and time scales of nanoseconds. Also, shear rates are usually some
orders of magnitude higher than in any
experiment~\cite{bib:lauga-brenner-stone-05}.
Due to the small accessible time and length scales of molecular dynamics
simulations, mesoscopic simulation methods such as the lattice Boltzmann method
are well applicable for the simulation of microfluidic experiments.

The experimental investigation of apparent slip can be based on different
setups: either a fluid is pumped through a microchannel and the measured
mass flow rate at the end of the channel is compared to the theoretical
value with no slip boundary conditions. From the deviation of the two
values, the magnitude of slip can be
computed~\REV{ED}{\cite{tretheway-meinhart-02}}. Another possibility is to measure
the slip length directly using optical methods like particle image
velocimetry (PIV). Very popular is
the modification of an atomic force microscope (AFM) by adding a silicon sphere
to the tip of the cantilever. While moving the sphere towards the
boundary, the required force is measured.  It is possible to measure the
amount of slip at the wall by comparing the force needed to move the
sphere with its theoretical
value~\cite{vinogradova-yakubov-03,bib:vinogradova-96}.

During the last few years, the substantial scientific research invested in
the slip phenomenon has lead to a more clear picture which can be
summarized as follows:
one can argue that many surprising results published were
only due to artefacts or misinterpretation of experiments. In general,
there seems to be an agreement within the community that slip lengths
larger than a few nanometers can usually be referred to as ``apparent slip''
and are often caused by experimental artefacts.
Small slip lengths are experimentally even harder to determine and require
sophisticated setups such as the modified atomic force microscopes as
described above. Here, small variations of the apparatus such as choosing
a different shape of the cantilever or modifying the control circuit of
the sample holder can lead to substantial variation of the measurements.
Also, the theoretical equations correlating the measured force to the slip
length are only valid for perfect surfaces and infinitely slow
oscillations of the sphere. Therefore, it is of importance to perform
computer simulations which have the advantage that most parameters can be
changed independently without modifying anything else. Thus, the influence
of every single modification can be studied in order to present estimates
of expected slip lengths.

\section{Apparent slip in hydrophobic microchannels}
The simulation method used to study microfluidic devices has to be chosen
carefully. While Navier-Stokes solvers are able to cover most problems in
fluid dynamics, they lack the possibility to include the influence of
molecular interactions as needed to model boundary slip. Molecular
dynamics simulations (MD) are the best choice to simulate the fluid-wall
interaction, but the computer power today is not sufficient to simulate
length and time scales necessary to achieve orders of magnitude which
are relevant for experiments. However, boundary slip with a slip length
$\beta$ of the order of many molecular diameters $\sigma$ has been studied
with molecular dynamics simulations by various authors~\REV{ED}{\cite{bib:thompson-troian-1997,bib:cieplak-koplik-banavar-01,bib:cottin-bizone-etall-04,baudry-charlaix-01,bib:priezjev-darhuber-troian-2005}}.

\REV{ED}{The current contribution focuses on numerical investigations of the
slip phenomenon by means of lattice Boltzmann simulations. While an emphasis is
put on reviewing our own contributions to the field, the achievements of other
groups are commonly referred to. However, it should be noticed that while a
large number of groups utilizes the lattice Boltzmann technique to investigate
microfluidic problems, only a very small number of researchers is actually
applying the method to studying slippage. Even though interactions have to be
described on a mesoscopic scale, this is surprising since mesoscopic simulation
methods offer a closer relation to experimentally relevant time and length
scales than microscopic techniques such as molecular dynamics.}

In the lattice Boltzmann method, one discretizes
the Boltzmann kinetic equation
\begin{equation}
\label{eq:boltzmann}
\left[\frac{\partial }{\partial t}+v\nabla_x\right] \eta({\bf x,v},t)={\bf \Omega}
\end{equation}
on a lattice. \REV{5}{The Boltzmann kinetic equation describes the evolution of the
single particle probability density $\eta({\bf x},{\bf v},t)$, where ${\bf x}$
is the position, ${\bf v}$ the velocity, and $t$ the time.}
The derivatives represent simple propagation of a single particle in real
and velocity space whereas the collision operator ${\bf \Omega}$ takes
into account molecular collisions in which a particle changes its momentum
due to a collision with another particle. To
represent the correct physics, the collision operator should conserve
mass and momentum, and should be Galilei invariant. By
performing a Chapman Enskog procedure, it can be shown that such a
collision operator ${\bf \Omega}$ reproduces the Navier-Stokes
equation~\cite{bib:succi-01}. In the lattice Boltzmann method the time $t$,
the position ${\bf x}$, and the velocity ${\bf v}$ are discretized.

A few groups have applied the lattice Boltzmann method for the simulation of
microflows and to study boundary slip. \REV{3+5}{A popular approach is to
introduce slip by generalizing the no-slip bounce back boundary conditions in
order to allow specular reflections with a given
probability~\cite{succi02,bib:tretheway-zhu-petzold-meinhart-2002,bib:tang-tao-he-2005,bib:sbragaglia-succi-2005}, or
to apply diffuse
scattering~\cite{bib:ansumaili-karlin-2002,bib:sofonea-sekerka-2005,bib:niu-shu-chew-2004}.
It has been shown by Guo et al. that these approaches are virtually
equivalent~\cite{bib:guo-shi-zhao-zheng-2007}.  Another possibility is to
modify the fluid's viscosity, i.e., the fluid viscosity is modified due to
local density variations in order to model slip~\cite{bib:nie-doolen-chen}. In
both cases, the parameters determining the properties at the boundaries are
``artificial'' parameters and they do not have any obvious physical meaning.
Therefore, they are not easily mappable to experimentally available values.  We
model the interaction between hydrophobic channel walls and the fluid by means
of a multi-phase lattice Boltzmann model. Our approach overcomes this problem
by applying a mesoscopic force between the walls and the fluid. A similar
approach is used by Zhu et al.~\cite{bib:zhu-tretheway-petzold-meinhart-2005},
Benzi et al.~\cite{bib:benzi-etal-06}, and Zhang et al.~\cite{bib:zhang-kwok-04}. This force applied at the boundary
can be linked to the contact angle which is commonly used by experimentalists
to quantitatively describe the wettability of a
material~\cite{benzi-etal-06b,bib:huang-thorne-schaap-sukop-2007}.}

The simulation method and our implementation of boundary conditions are
described as follows. A multiphase lattice Boltzmann system can be
represented by a set of equations
\begin{equation}
\label{LBeqs}
\begin{array}{cc}
\eta_i^{\alpha}({\bf x}+{\bf c}_i, t+1) - \eta_i^{\alpha}({\bf x},t) =
\Omega_i^{\alpha}, &  i= 0,1,\dots,b\mbox{ ,}
\end{array}
\end{equation}
where $\eta_i^{\alpha}({\bf x},t)$ is the single-particle distribution
function, indicating the amount of species $\alpha$ with velocity ${\bf
c}_i$, at site ${\bf x}$ on a D-dimensional lattice of coordination number
$b$ (D3Q19 in our implementation), at time-step $t$. This is a discretized
version of equation (\ref{eq:boltzmann}) without external forces ${\bf F}$
for a number of species $\alpha$. For the collision operator
$\Omega_i^{\alpha}$ we choose the Bhatnagar-Gross-Krook (BGK)
form~\cite{bib:bgk}
\begin{equation}
\label{Omega}
 \Omega_i^{\alpha} =
 -\frac{1}{\tau^{\alpha}}(\eta_i^{\alpha}({\bf x},t) - \eta_i^{\alpha
\, eq}({\bf u}^{\alpha}({\bf x},t),\eta^{\alpha}({\bf x},t)))\mbox{ ,}
\end{equation}
where $\tau^{\alpha}$ is the mean collision time for component $\alpha$
and determines the kinematic viscosity
\begin{equation}
\label{eq:vis}
\nu^\alpha=\frac{2\tau^\alpha-1}{6} . 
\end{equation}
of the fluid. \REV{1}{The
relaxation time $\tau^\alpha$ is kept constant at $1.0$ in this study.}
The system relaxes to an equilibrium distribution
$\eta_i^{\alpha\,eq}$ which can be derived imposing restrictions on the
microscopic processes, such as explicit mass and momentum conservation for
each species.
In our implementation we choose for the equilibrium distribution function
\begin{equation}
\begin{array}{ll}
\label{eq:equil}
\eta_i ^{eq} =\\
\zeta_i\eta^{\alpha}
\left[\! 1+
\frac{{\bf c}_i \cdot {\bf u}}{c_s^2}\! +\!\frac{({\bf c}_i \cdot {\bf
u})^2}{2c_s^4}
- \frac{u^2}{2c_s^2} + \frac{({\bf c}_i \cdot {\bf u})^3}{6c_s^6}
- \frac{u^2({\bf c}_i \cdot {\bf u})}{2c_s^4}\right]\! ,
\end{array}
\end{equation}
which is a polynomial expansion of the Maxwell distribution.
$c_i$ are the velocity vectors pointing to neighbouring lattice sites \REV{1}{and $\zeta_i$ are the lattice weights resulting from the velocity space discretization}.
$c_s=1/\sqrt{3}$ is the speed of sound for the D3Q19 lattice.
The macroscopic values can be derived from the single-particle
distribution function $\eta^{\alpha}_i({\bf x},t)$, i.e., the density
$\eta^{\alpha}({\bf x},t)$ of the species $\alpha$ at lattice site ${\bf x}$
is the
sum over the distribution
functions $\eta^{\alpha}_i({\bf x},t)$ for all lattice velocities
${\bf c_i}$,
\begin{equation}
\eta^{\alpha}({\bf x},t)\equiv\sum_i \eta_i^{\alpha}({\bf x},t).
\end{equation}
${\bf u}^{\alpha}({\bf x},t)$ is the macroscopic
velocity of the fluid, defined as
\begin{equation}
\eta^{\alpha}({\bf x},t){\bf u}^{\alpha}({\bf x},t) \equiv \sum_i
\eta_i^{\alpha}({\bf x},t){\bf c}_i.
\end{equation}
Interactions between different fluid species are introduced following Shan
and Chen as a mean field body force between nearest neighbors
\cite{bib:shan-chen-93,bib:shan-chen-liq-gas},
\begin{equation}
\label{Eq:force}
{\bf F}^{\alpha}({\bf x},t) \equiv -\psi^{\alpha}({\bf x},t)\sum_{\bf
\bar{\alpha}}g_{\alpha \bar{\alpha}}\sum_{\bf 
x^{\prime}}\psi^{\bar{\alpha}}({\bf x^{\prime}},t)({\bf x^{\prime}}-{\bf
x})\mbox{ ,}
\end{equation}
where $\psi^{\alpha}({\bf x},t)=(1 - e^{-\eta^{\alpha}({\bf
x},t)/\eta_0})$ is the so-called effective mass with $\eta_0$ being a
reference density that is set to $1$ in our case~\cite{bib:shan-chen-93}.
$g_{{\alpha}\bar{\alpha}}$ is a force coupling constant, whose magnitude
controls the strength of the interaction between component $\alpha$ and
$\bar{\alpha}$. The dynamical effect of the force is realized in the BGK
collision operator (\ref{Omega}) by adding an increment $\delta{\bf
u}^{\alpha} = {\tau^{\alpha}{\bf F}^{\alpha}}/{\eta^{\alpha}}$ to the velocity
${\bf u}$ in the equilibrium distribution function (\ref{eq:equil}). \REV{1}{A
repulsive potential between surface and fluid can be used to model hydrophobic
fluid-surface interactions. Such a potential is realized by attaching the
imaginary fluid ``density'' $\eta^{\rm wall}$ to the first lattice site inside
the wall. Only the distribution corresponding to the rest velocity is filled,
while the remaining ones are kept at 0. As a result the only difference between
$\eta^{\rm wall}$ and any other fluid packages on the lattice
$\eta^{\bar{\alpha}}$ is that the fluid corresponding to $\eta^{\rm wall}$ is
only taken into account for in the collision step and for the calculation of
Eq.~\ref{Eq:force}, but not in the propagation step. Therefore, we can adopt
$\eta^{\rm wall}$ and the coupling constant $g_{\alpha, \rm wall}$ in order to
tune the fluid-wall interaction.} $g_{\alpha, \rm wall}$ is kept at $0.08$
throughout this paper if not mentioned otherwise and all values are reported in
lattice units. These parameters allow to simulate a wide range of effective
interactions without compromising on numerical stability. \REV{2}{Additionally, we
apply second order correct mid-grid bounce back boundary conditions between the
fluid and the surface which assures vanishing velocities at solid surfaces. Here,
a distribution function that would be advected into a solid node is simply
reversed and advected into the opposite direction~\cite{bib:succi-01}.}

From molecular dynamics simulations it is known that the fluid-wall
interactions causing a slip phenomenon usually take place within a few
molecular layers of the liquid along the boundary
surface~\cite{bib:thompson-troian-1997,bib:cieplak-koplik-banavar-01,bib:cottin-bizone-etall-04,baudry-charlaix-01}.
Our coarse-grained fluid wall interaction acts on the length scale of one
lattice constant and does not take the molecular details into account.
Therefore, \REV{ED}{coarse-grained implementations based on the lattice
Boltzmann method} are only able to reproduce an averaged effect of the
interaction and cannot fully resolve the correct flow profile very close to the
wall and below the resolution of a single lattice spacing.
However, in order to understand the influence of
the hydrophobicity on experimentally observed apparent slip, it is fully
sufficient to investigate the flow behavior on more macroscopic scales as
they are accessible for experimental investigation. \REV{ED}{Coarse-grained interaction models could be
improved by a direct mapping of data obtained from MD simulations to the
coupling constant $g_{\alpha, \rm wall}$ allowing a direct comparison of
the influence of liquid-wall interactions on the detected
slip~\cite{bib:jens-kunert-herrmann:2005}. Similar approaches are known from quantitative comparisons of lattice Boltzmann and molecular dynamics simulations in the literature~\cite{bib:horbach-suci-2006,bib:chibbaro-etal-2008}.}

The simulations in this work use a setup of two infinite planes separated
by the distance $2d$. We call the direction between the two planes $x$ and
if not stated otherwise $2d$ is set to $64$ lattice sites. In $y$
direction we apply periodic boundary conditions. Here, $8$ lattice sites
are sufficient to avoid finite size effects since there is no propagation
in this direction. $z$ is the direction of the flow with our channels
being $512$ lattice sites long. At the beginning of the simulation ($t=0$)
the fluid is at rest. We then apply a pressure gradient $\nabla P$ in the
$z$- direction to generate a planar Poiseuille flow. Assuming Navier's
boundary condition, the slip length $\beta$ is measured by fitting the
theoretical velocity profile as given by equation \ref{eq:plattenprofil}
in flow direction ($v_z$) at position $x$, to the simulated data via the
slip length $\beta$. We validate this approach by comparing the measured
mass flow rate $\int \eta v(x) {\rm dx}$ to the theoretical mass flow
without boundary slip and find a very good agreement. The dynamic
viscosity $\mu$ as well as the pressure gradient $\frac{\partial
P}{\partial z}$ needed to fit equation (\ref{eq:plattenprofil}) are
obtained from our simulation data.

In~\cite{bib:jens-kunert-herrmann:2005}, we show that
this model creates a larger slip $\beta$ with stronger interaction, namely
larger $g_{\alpha,\rm wall}$ and larger $\eta^{\rm wall}$. The maximum
available slip length measured is $5.0$ in lattice units. For stronger
repulsive potentials, the density gradient at the fluid-wall interface becomes
too large, causing the simulation to become unstable. At lower interactions the
method is very stable and the slip length $\beta$ is independent of the
distance $d$ between the two plates and therefore independent of the
resolution. We also show that the slip decreases with increasing pressure since
the relative strength of the repulsive potential compared to the bulk pressure
is weaker at high pressure.  Therefore, the pressure reduction near the wall is
less in the high pressure case than in the low pressure one. Furthermore, we
demonstrate that $\beta$ can be fitted with a semi analytical model based on a
two viscosity model.

\begin{figure}[h]
\centerline{
\centerline{\includegraphics[height=5cm]{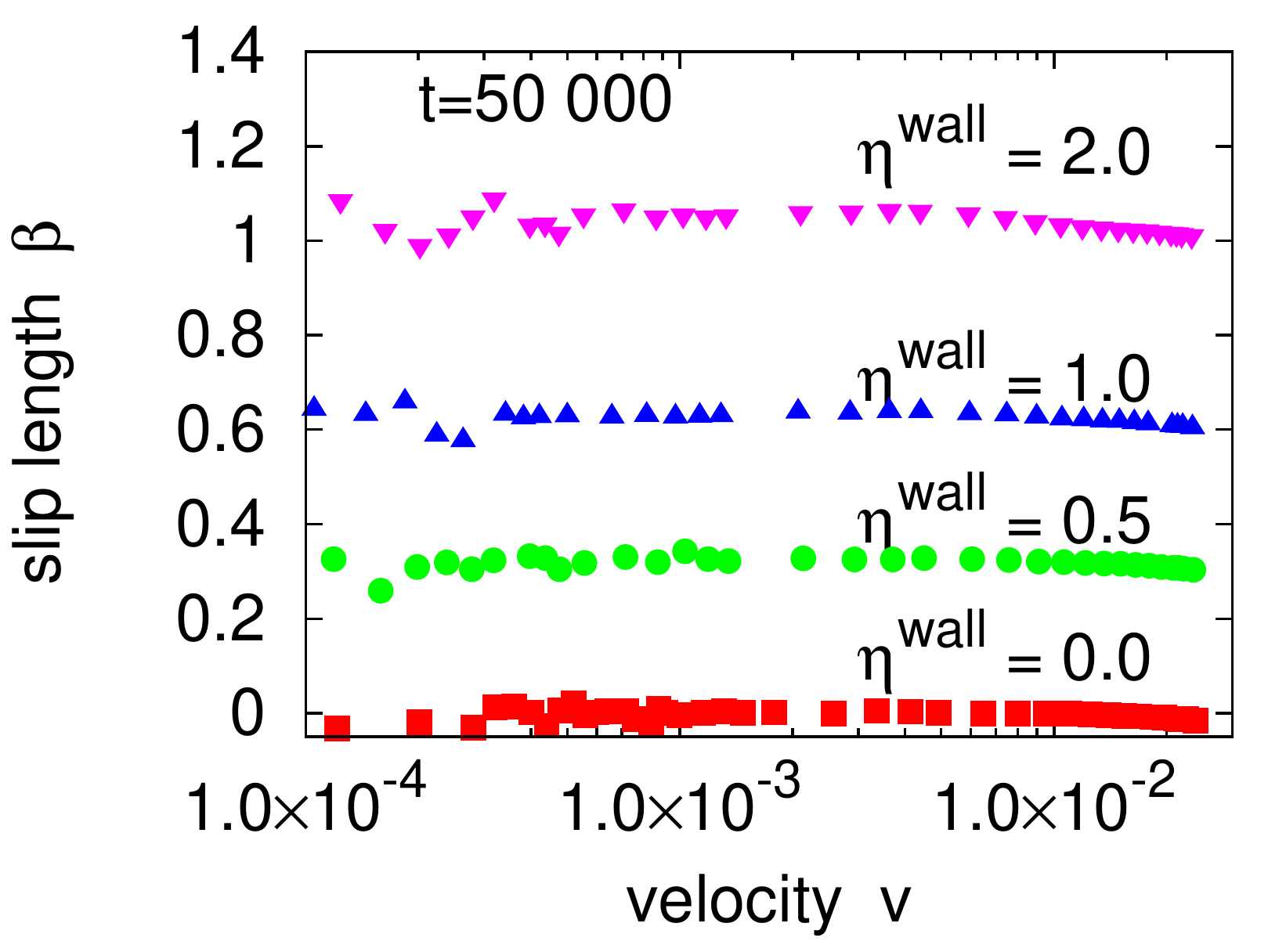}}
}
\caption{\label{fig:slipvsvel}
Slip length $\beta$ versus bulk velocity $v$ for different fluid-wall
interactions $\eta^{\rm wall}$. $\beta$ is independent of $v$ and only
depends on $\eta^{\rm wall}$~\cite{bib:jens-kunert-herrmann:2005}. All units are
expressed in lattice units throughout this paper if not stated otherwise.}
\end{figure}
We study the dependence of the slip length $\beta$ on the flow velocity for a
wide range of velocities of more than three decades as it can be seen in
Fig.~\ref{fig:slipvsvel} and in~\cite{bib:jens-kunert-herrmann:2005}. In the
figure, we show data for different fluid-wall interactions $0 <\eta^{wall}<2.0$
and flow velocities from $10^{-4}<v<10^{-1}$. \REV{1}{For simplicity we
restrict ourselves to $g_{\alpha,\rm wall}=0.08$ which is a suitable value found from parameter studies given in~\cite{bib:jens-kunert-herrmann:2005}.} Within this region we confirm
the findings of many steady state experiments~\cite{cheng-giordano-02}, i.e.,
that the slip length is independent of the flow velocity and only depends on
the wettability of the channel walls.  Some dynamic
experiments, however, find a shear rate dependent
slip~\cite{neto-craig-williams-03,bib:zhu-granick-01}.
These experiments often utilize a modified AFM as
described in the introduction to detect boundary slippage. Since the slip
length is found to be constant in our simulations after sufficiently long
simulation times, we cannot confirm these results.  However, it has been
proposed by various authors that this velocity dependence is due to
non-controlled effects such as impurities or surface nanobubbles.
\REV{3}{In simulations we can only find a shear rate dependence if the system
has not yet reached the steady state or if time-dependent accelerations are
present~\cite{bib:jens-kunert:2008b}.}

Our mesoscopic approach is able to reach the small flow velocities of
known experiments and reproduces results from experiments and other
computer simulations, namely an increase of the slip with increasing
liquid-solid interactions, the slip being independent of the flow
velocity, and a decreasing slip with increasing bulk pressure. In
addition, within our model we develop a semi-analytic approximation of the
dependence of the slip on the bulk pressure as described
in~\cite{bib:jens-kunert-herrmann:2005}.

\section{Roughness induced apparent slip}
\begin{figure}[h]
\centerline{\includegraphics[height=3.2cm]{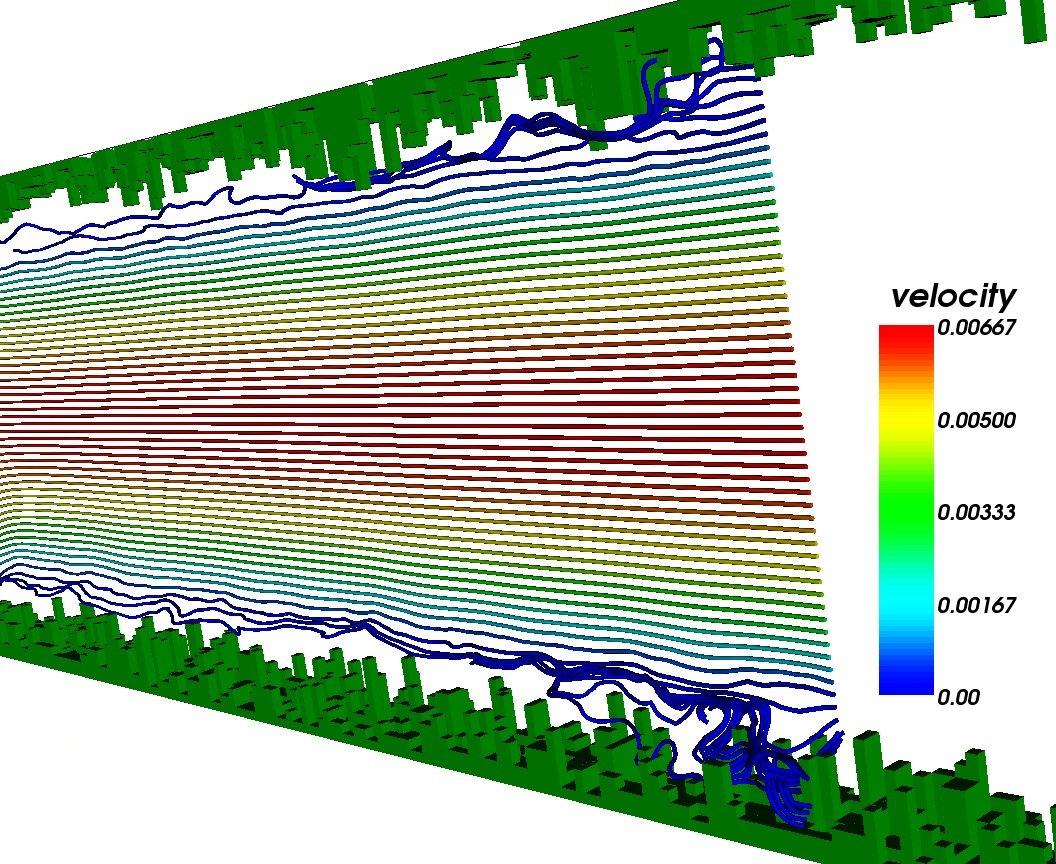}\quad\includegraphics[height=3.2cm]{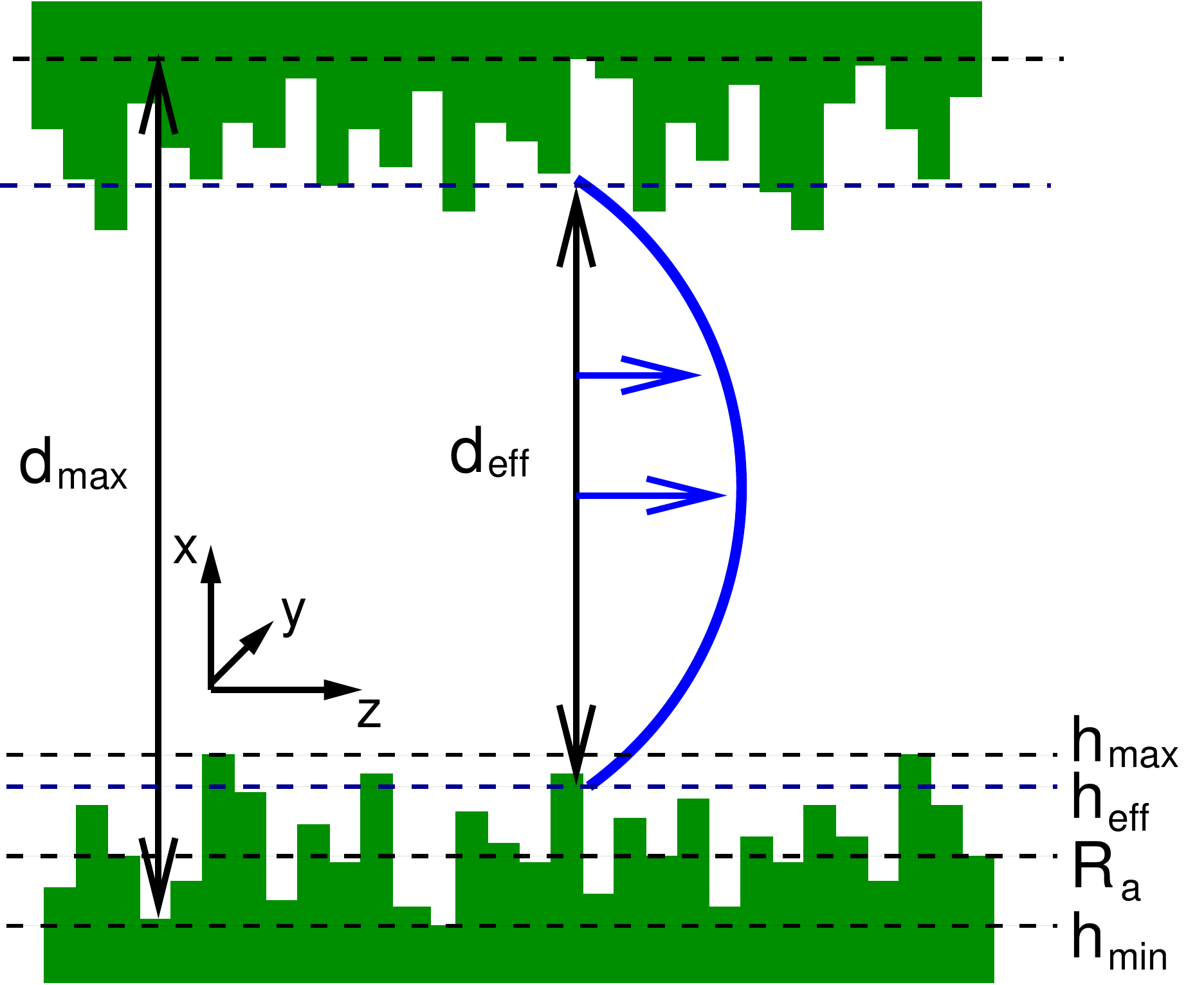}}
\caption{\label{fig:zufall}Left: a typical simulated system. Poiseuille flow
between two rough surfaces showing random surface variations. Streamlines
depict a two dimensional cut and illustrate the parabolic velocity
profile. This profile is distorted in the vicinity
of the rough surfaces~\cite{bib:jens-kunert:2008c}. \REV{1}{Right: the effective boundary height
$h_{\rm eff}$ is found between the deepest valley at $h_{\rm min}$ and
the highest peak at $h_{\rm max}$. It corresponds to an effective
channel width $d_{\rm eff}$. $R_a$ denotes the average roughness and the maximum distance between the
plates $d_{\rm max}$ is kept constant~\cite{bib:jens-kunert:2008c}.}}
\end{figure}

If typical length scales of the experimental system are comparable to the
scale of surface roughness, the effect of roughness cannot be neglected
anymore.
Figure~\ref{fig:zufall}(left) shows a typical example of a simulation setup:
Poiseuille flow between two rough surfaces. 
\REV{2}{The surface is generated using a random number generator to randomly choose the height of the obstacles at every discrete surface position.}
As can be observed in the
figure, the stream lines of the flow are getting disturbed or trapped
between the obstacles at the surfaces. In this section we show that
an apparent boundary slip can have its origin in the misleading assumption
of perfectly smooth boundaries.

The influence of surface variations on the slip length $\beta$ has been
investigated by numerous authors. \REV{1,ED}{It was de\-mon\-stra\-ted by Richardson that roughness leads to higher
drag forces and thus to no-slip on macroscopic scales. He has shown that if on a rough surface even a
full-slip boundary condition is applied, one obtains a flow speed reduction
near the boundary resulting in a macroscopic no-slip boundary
condition~\cite{bib:richardson-73}. An experimental confirmation was later
presented by McHale and Newton~\cite{mchale-newton-04}. Molecular dynamics
(MD) simulations of Couette flow between sinusoidal walls have been presented
by Jabbarzadeh et al.~\cite{jabbarzadeh-etal-00}. They found that slip appears
for roughness amplitudes smaller than the molecular length
scale~\cite{jabbarzadeh-etal-00}. Sbragaglia et al. applied the LB method to
simulate fluids in the vicinity of microstructured hydrophobic
surfaces~\cite{bib:sbragaglia-etal-06}, Al-Zoubi et al. demonstrated that the
LB method is well applicable to reproduce known flow patterns in sinusoidal
channels~\cite{bib:alzoubi-brenner-2008}, and Varnik et
al.~\cite{varnik-raabe-06,varnik-dorner-raabe-06} have shown that even in small
geometries rough channel surfaces can cause flow to become turbulent.} 

Recently, we presented the idea of an effective wall for rough channel
surfaces~\cite{bib:jens-kunert:2007b}. Here, we investigate the influence of
different types of roughness on the position of the effective boundary.
Further, we show how the effective boundary depends on the distribution of the
roughness elements and how roughness and hydrophobicity interact with each
other~\cite{bib:jens-kunert:2008c}. Lecoq and coworkers~\cite{lecoq-etal-04}
performed experiments with well defined roughness, and developed a theory to
predict the position of the effective boundary. In the experiments they
utilised a laser interferometer to measure the trajectory of a colloidal
sphere, and thereby determined the lubrication force and an effective boundary
position. The used geometry consists of grooves with a triangular profile. For
a theoretical description the boundary is expressed in a Fourier series that
gives the boundary condition for the Laplace equation. From this an effective
boundary can be derived by a fast converting series.

In this paper, we revise our previous achievements and compare them with the
theoretical and experimental results of Lecoq and
coworkers~\cite{lecoq-etal-04}.

Again, Poiseuille flow measurements are utilized to investigate the effect
of interest. The rough surfaces are characterised by the highest point of
one plane ($h_{\rm max}$), the position of the deepest valley ($h_{\rm
min}$) and the arithmetic average of all surface heights giving the
average roughness $R_a$. In the case of symmetrical distributions we get
$R_a=h_{\rm max}/2$.

The position of the effective boundary $h_{\rm eff}$ can be found by
fitting the parabolic flow profile via the distance $d_{\rm eff}$.
With $\beta$ set to 0 we obtain the no-slip case.
To obtain an average value for the effective distance between the planes
$d_{\rm eff}$, a sufficient number of individual profiles at different
positions $z$ are taken into account. The so found $d_{\rm eff}$ gives
the position of the effective boundary and the effective height $h_{\rm
eff}$ of the rough surface is then defined by $d_{\rm max}-d_{\rm eff}$ (see
Fig.~\ref{fig:zufall}, left).

\begin{figure}[h]
\centerline{\includegraphics[height=5cm]{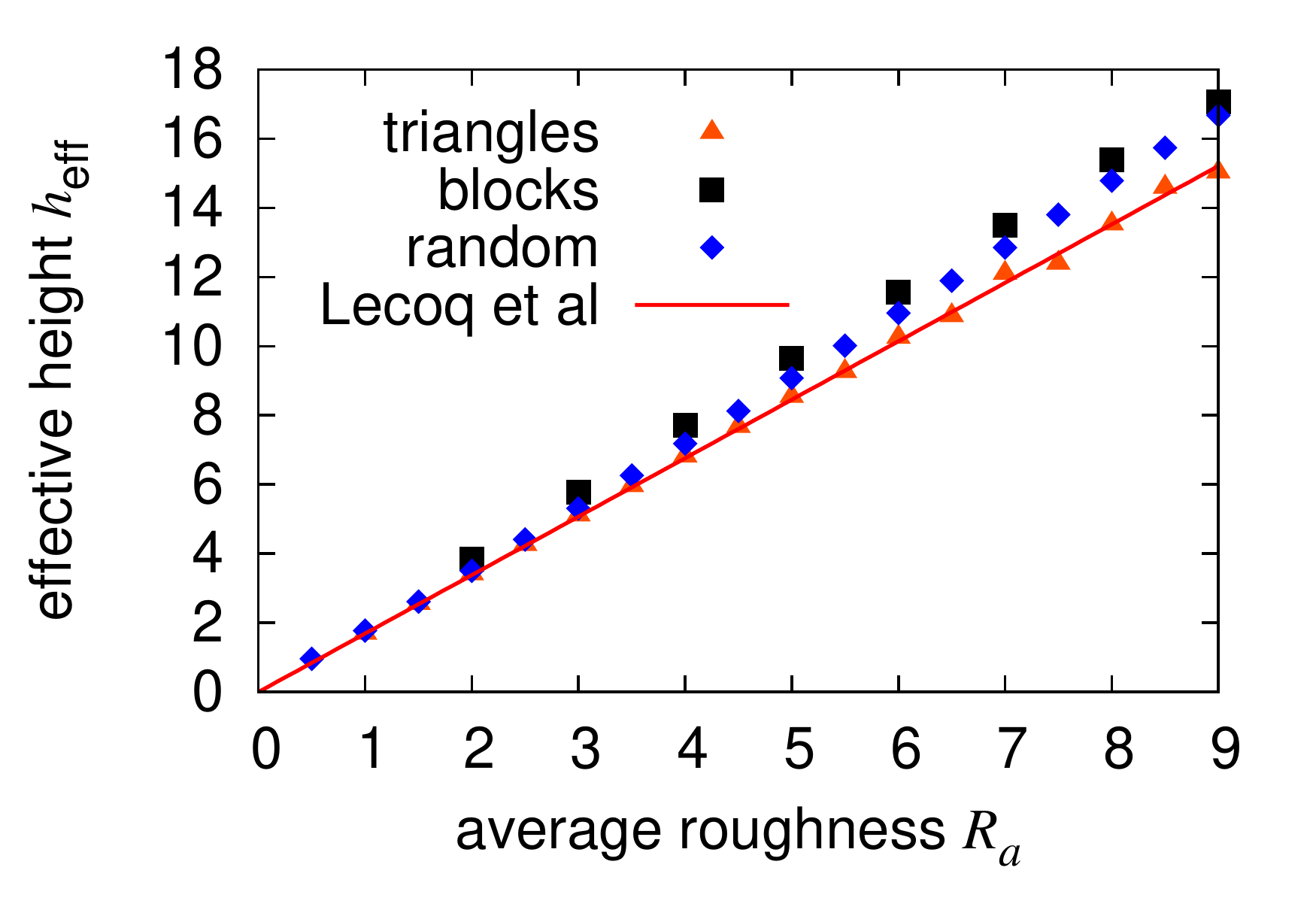}}
\caption{ \label{fig:lecoq}
Simulated effective height $h_{\rm eff}$ versus $R_a$ for
different surface geometries. The triangular shape matches the theoretical
results of Lecoq et al.~\cite{lecoq-etal-04} for a similar geometry.
}
\end{figure}

We show that the position of the effective boundary height is depending
on the shape of the roughness elements, i.e., for strong surface
distortions it is between $1.69$ and $1.90$ times the average height of
the roughness $R_a=h_{\rm max}/2$~\cite{bib:jens-kunert:2007b}. 
In Fig~\ref{fig:lecoq} we plot the effective boundary positions of
different geometries, i.e. randomly distributed grooves with a square
profile and grooves with a triangular profile. The results for the
triangular ones
match with the theoretical value of Lecoq et al.~\cite{lecoq-etal-04}
for a similar geometry.

By adding an additional distance between roughness elements, $h_{\rm eff}$
decreases slowly, so that the maximum height is still the leading parameter. We
are also able to simulate flow over surfaces generated from AFM data of gold
coated glass used in microflow experiments by O.I. Vinogradova and G.E.
Yakubov~\cite{vinogradova-yakubov-06}. We find that the height distribution of
such a surface is Gaussian and that a randomly arranged surface with a similar
distribution gives the same result for the position of the effective boundary
although in this case the heights are not correlated.
\begin{figure}[h]
\centerline{\includegraphics[height=5cm]{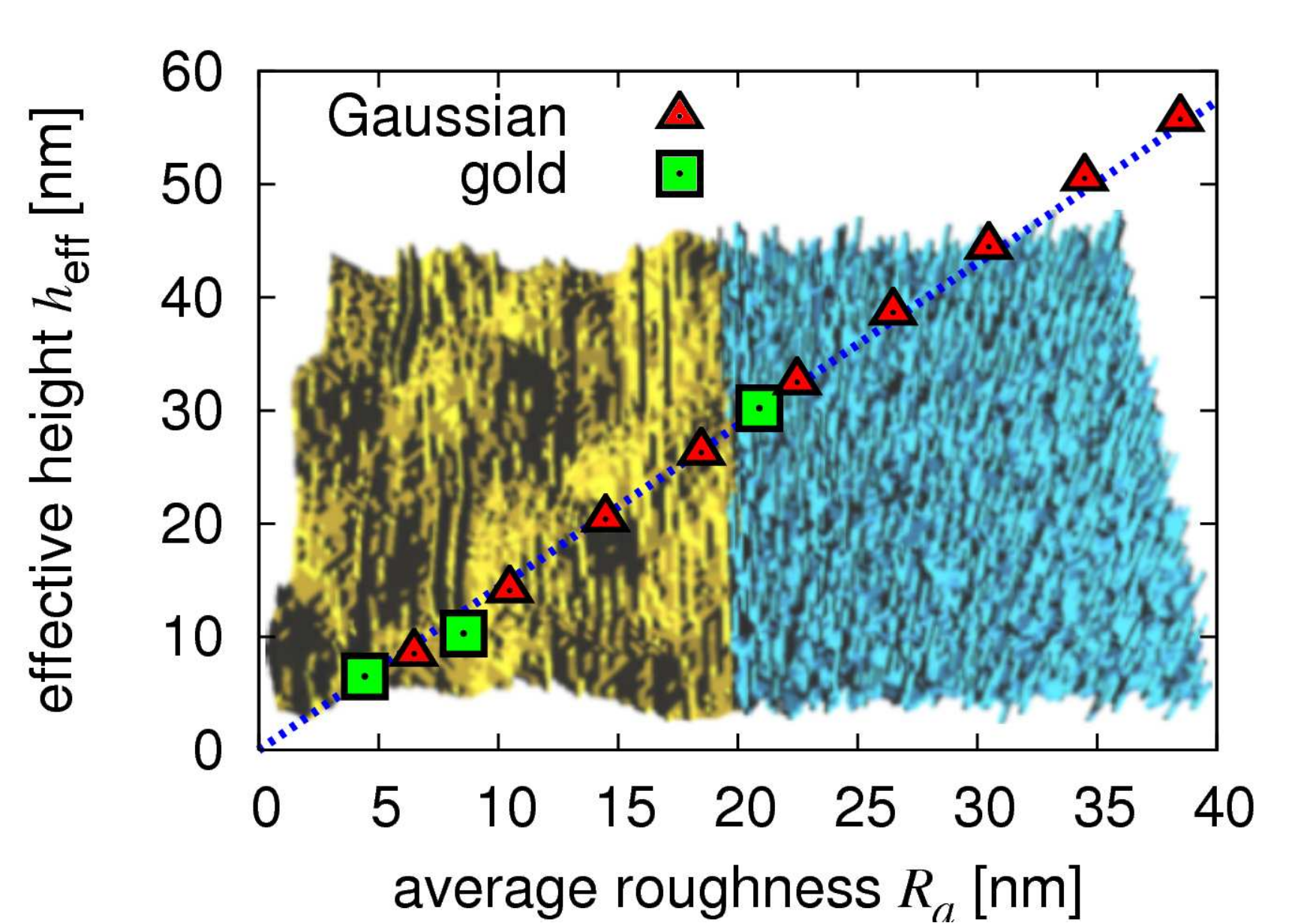}}
\caption{ \label{fig:afmrough}
Simulated effective height $h_{\rm eff}$ versus $R_a$ for gold
coated glass surfaces and a randomly generated surface with Gaussian
distributed heights. The background image shows the gold surface on the
left and the artificially generated structure on the right~\cite{bib:jens-kunert:2007b}.
}
\end{figure}

We can tune the width of the distribution $\sigma$ and the average height
$R_a$. By scaling $\sigma$ with $R_a$ we obtain geometrically similar
geometries. This similarity is important because the effective height $h_{\rm
eff}$ scales with the average roughness in the case of geometrical
similarity~\cite{bib:jens-kunert:2007b}. We investigate Gaussian distributed
heights with different widths $\sigma$ and find that the height of the
effective wall depends linearly on $\sigma$ in the observed
range~\cite{bib:jens-kunert:2008c}.  Further, we find that the slip diverges as
the amplitude of the roughness increases and the flow field gets more
restricted which highlights the importance of a proper treatment of surface
variations in very confined geometries~\cite{bib:jens-kunert:2007b}.

\section{Structured surfaces with entrapped microbubbles}
A natural continuation of our previous works on roughness induced apparent
boundary slip and the collaboration mentioned above is the analysis of flow
along superhydrophobic surfaces~\cite{bib:jens-jari:2008}. While in typical
experiments, slip lengths of a few tens of nanometers can be observed, it would
be preferable for technical applications to increase the throughput of fluid in
a microchannel, i.e., to obtain substantially larger slip. Superhydrophobic
surfaces are promising in this context, since it has been recently
predicted~\cite{bib:cottin-bizone-etall-03} and experimentally
reported~\cite{bib:perot-rothstein-04} that the so-called Fakir effect or
Cassie state considerably amplifies boundary slippage. Using highly rough
hydrophobic surfaces such a situation can be achieved. Instead of entering the
area between the rough surface elements, the liquid remains at the top of the
roughness and traps air in the interstices. Thus, a very small liquid-solid
contact area is generated.

Steinberger et al. utilized surfaces patterned with a square array of
cylindrical holes to demonstrate that gas bubbles present in the holes may
cause a reduced
slip~\cite{bib:steinberger-cottin-bizonne-kleimann-charlaix:2007}.
Numerically, they found even negative slip lengths for flow over such bubble
mattresses, i.e., the effective no-slip plane is inside the channel and the
bubbles increase the flow resistance.  In this section we consider negative
slip lengths on bubble surfaces and also discuss the question of shear-rate
dependent slip. In particular we show that microbubbles can generate a
shear-rate dependence.

Our simulations utilize the single component multiphase LB model by Shan
and Chen~\cite{bib:shan-chen-liq-gas} which enables simulations of
liquid-vapor systems with surface tension. 
\REV{ED}{We are not aware of further lattice Boltzmann simulations to study the flow over a bubble mattress. 
However, a number of authors has applied various LB multiphase and
multicomponent models to study the properties of droplets on chemically
patterned and superhydrophobic
surfaces~\cite{kusumaatmaja-06,bib:kusumaatmaja-yeomans-2007,bib:pirat-etal-2008,bib:hyvaluoma-koponen-raiskinmaki-timonen:2007}.}
The flow in our system is confined between
two parallel walls. One of the walls is patterned with holes and vapor
bubbles are trapped to these holes. The other wall is smooth and moved
with velocity $u_0$. Steinberger et
al.~\cite{bib:steinberger-cottin-bizonne-kleimann-charlaix:2007} presented
finite-element simulations of flow over rigid ``bubbles'' by applying slip
boundaries at static bubble surfaces. The LB method allows the bubbles to
deform if the viscous forces are high enough compared to the surface
tension. We are also interested in how surface patterning affects the slip
properties of these surfaces, and how bubbles could be utilized to develop
surfaces with special properties for microfluidic
applications~\cite{bib:jens-jari:2008}. 

\begin{figure}[h]
\centerline{\includegraphics[height=3.8cm]{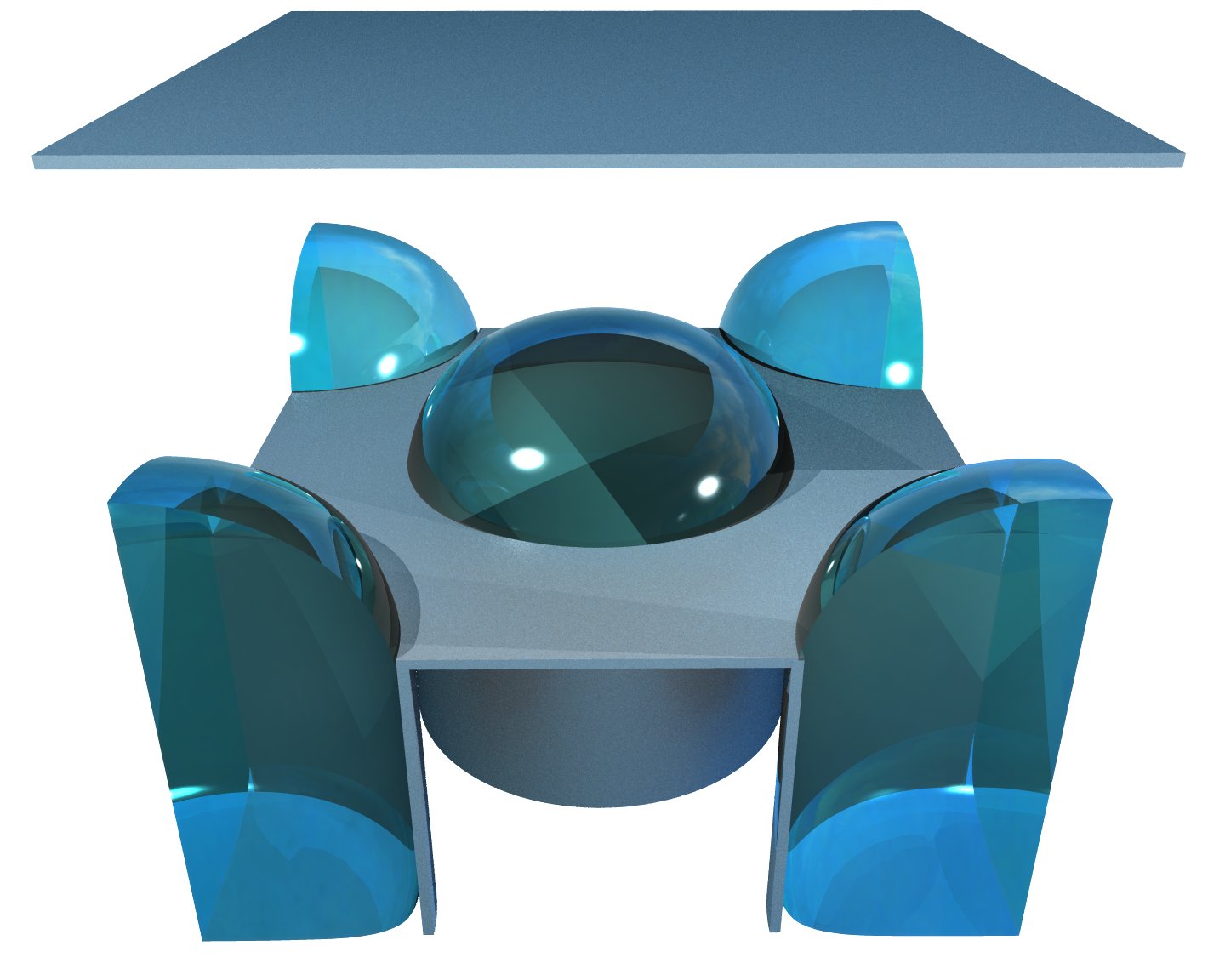}\includegraphics[height=4cm]{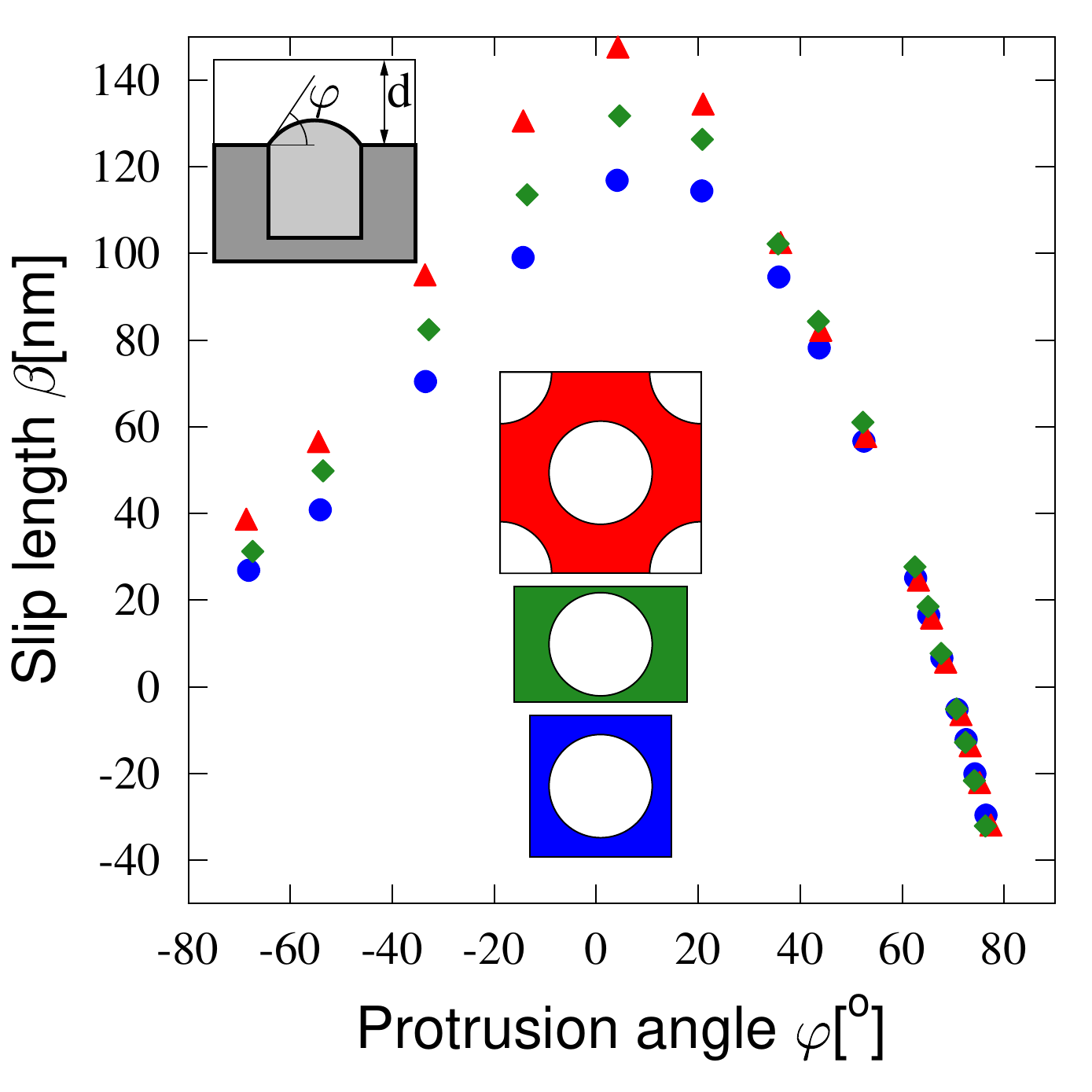}}
\caption{\label{fig:slip}
A visualisation of the simulation setup (left): the lower surface is
patterned with holes, while the upper surface is moved with
velocity $u_0$.
Right: the slip length $\beta$ as a function of protrusion angle
$\varphi$.
A unit cell of each array is shown in insets and corresponding
results are given by triangles (rhombic array), diamonds (rectangular
array), and circles (square array). The inset in the top-left corner
shows the definition of $\varphi$~\cite{bib:jens-jari:2008}.}
\end{figure}

The distance between walls is $d=1$ $\mu$m (40 lattice nodes) in all
simulations, and the area fraction of holes is $0.43$.  A unit cell of the
regular array is included in a simulation and periodic boundary condition are
applied at domain boundaries. The bubbles are trapped to holes by using
different wettabilities for boundaries in contact with the main channel and
with the hole. The protrusion angle $\varphi$ (see Fig.~\ref{fig:slip} for
definition) is varied by changing the liquid's bulk pressure.  The effective
slip length is $\beta = \mu u_0/ \sigma - d$, where  $\sigma = \mu {\rm
d}v/{\rm d}z$ is the shear stress acting on the upper wall and $\mu$ the
dynamic viscosity of the liquid.

We investigate
the effect of a modified protrusion angle and different surface
patterns by using square, rectangular, and rhombic bubble arrays. The
cylindrical holes have a radius $a = 500$~nm and the area fraction of the holes
is equal in all cases. The shear rate is such that the Capillary number $Ca =
\mu a G_s /\gamma = 0.16$. Here, $G_s$ and $\gamma$ are the shear rate and
surface tension, respectively. A snapshot of a simulation is shown in the left
part of Fig.~\ref{fig:slip} and the slip lengths obtained are shown in the
right part. The observed behavior is similar to that reported
in~\cite{bib:steinberger-cottin-bizonne-kleimann-charlaix:2007}, where a square
array of holes was studied. In particular, we observe that when $\varphi$ is
large enough $\beta$ becomes negative. Moreover, when the protrusion angle
equals zero, the slip length is maximised and the highest possible throughput
in a microchannel is obtained. The behavior of the slip length can be explained
by thinking of an increased surface roughness if the protrusion angle is larger
or smaller than zero.  Since the area fraction of the bubbles is the same in
all three cases, our results clearly indicate that slip properties of the
surface can be tailored not only by changing the protrusion angle but also by
the array geometry. In the presented study, the highest slip lengths are
obtained for the rhombic unit cell and it is current work of progress to
investigate the influence of the array geometry in more detail.
\REV{ED}{Recently, our findings have been confirmed theoretically by Davis and
Lauga~\cite{bib:davis-lauga-2009}.}

\begin{figure*}[h]
\centerline{\includegraphics[height=3.8cm]{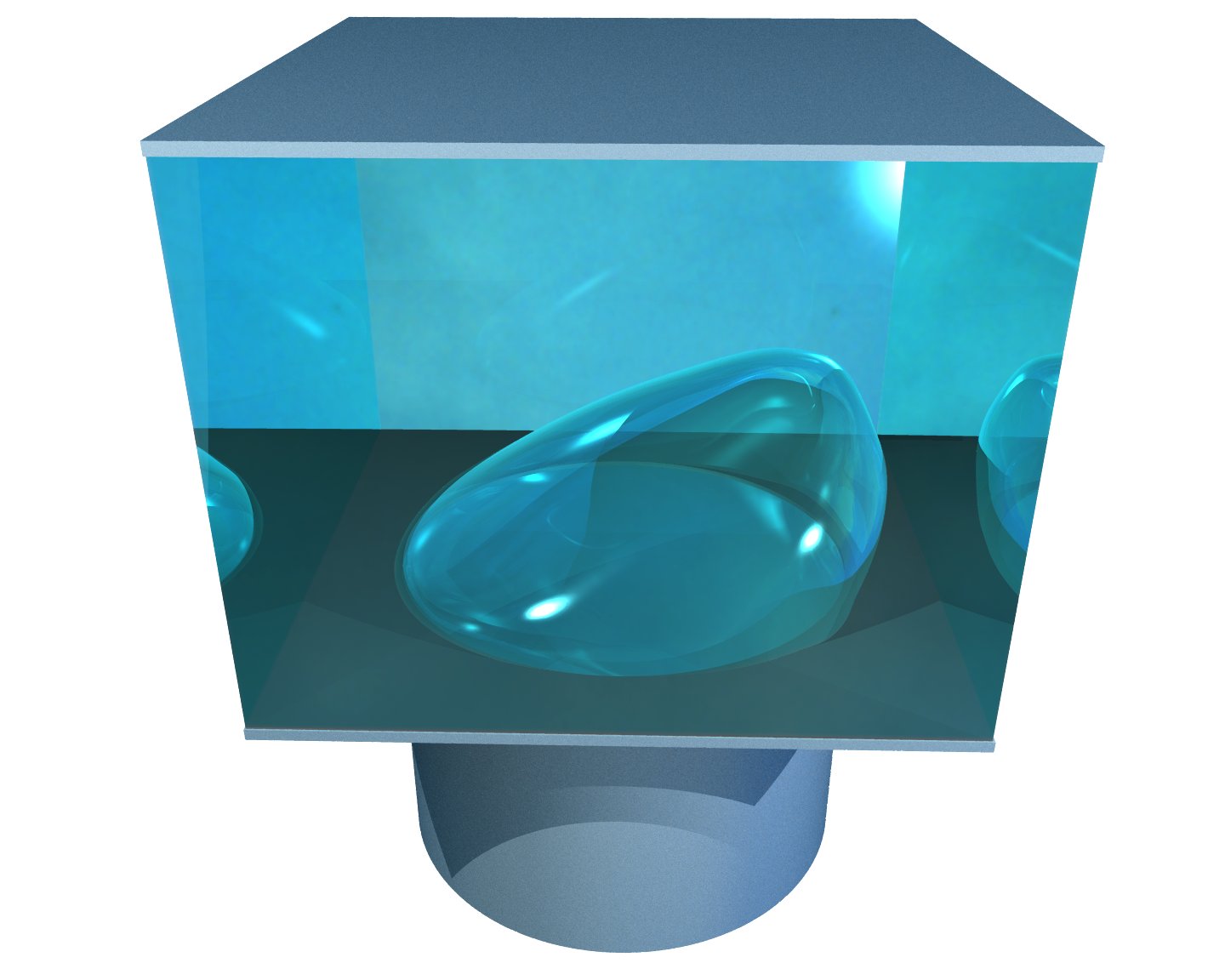}\includegraphics[height=3.8cm]{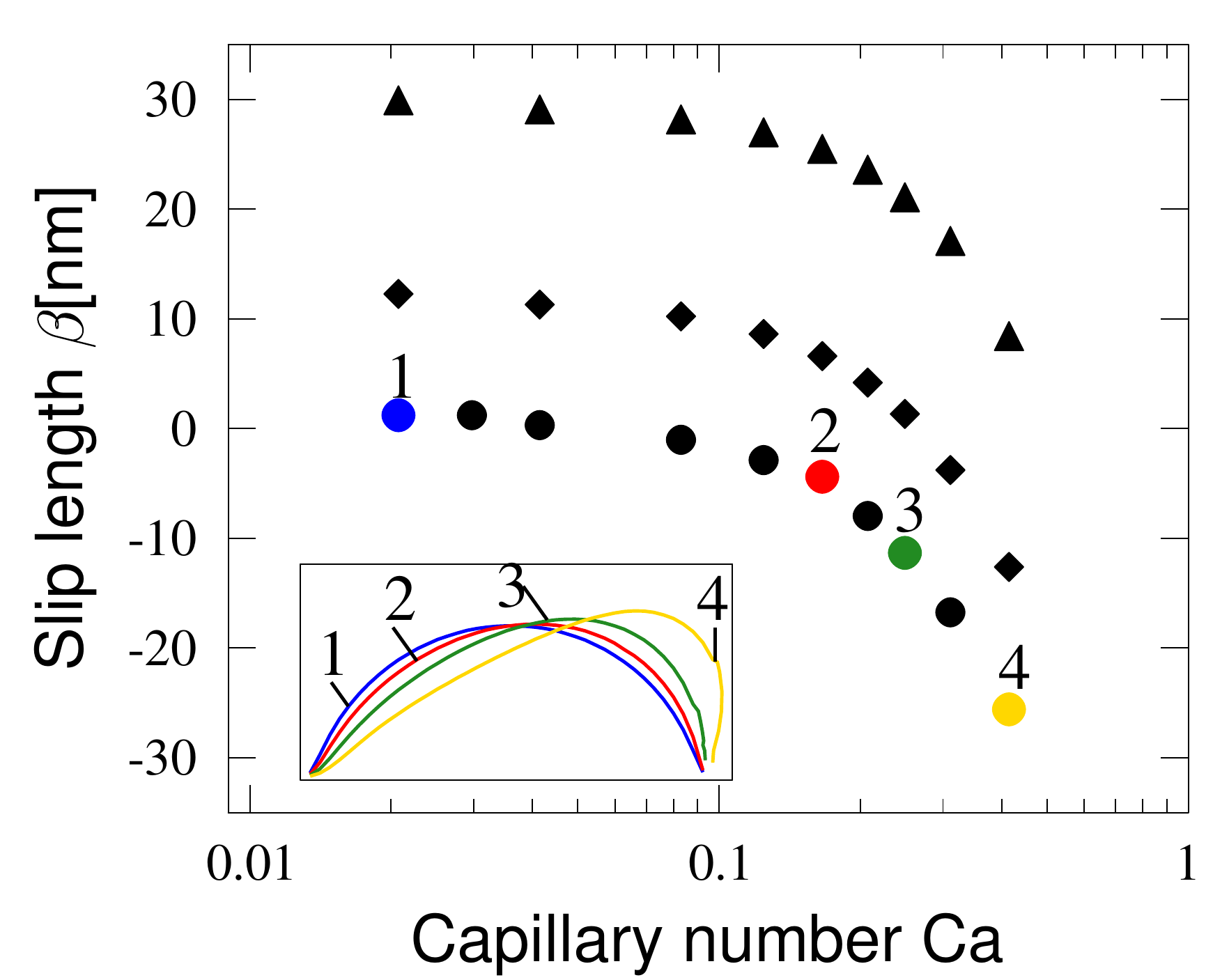}\includegraphics[height=3.8cm]{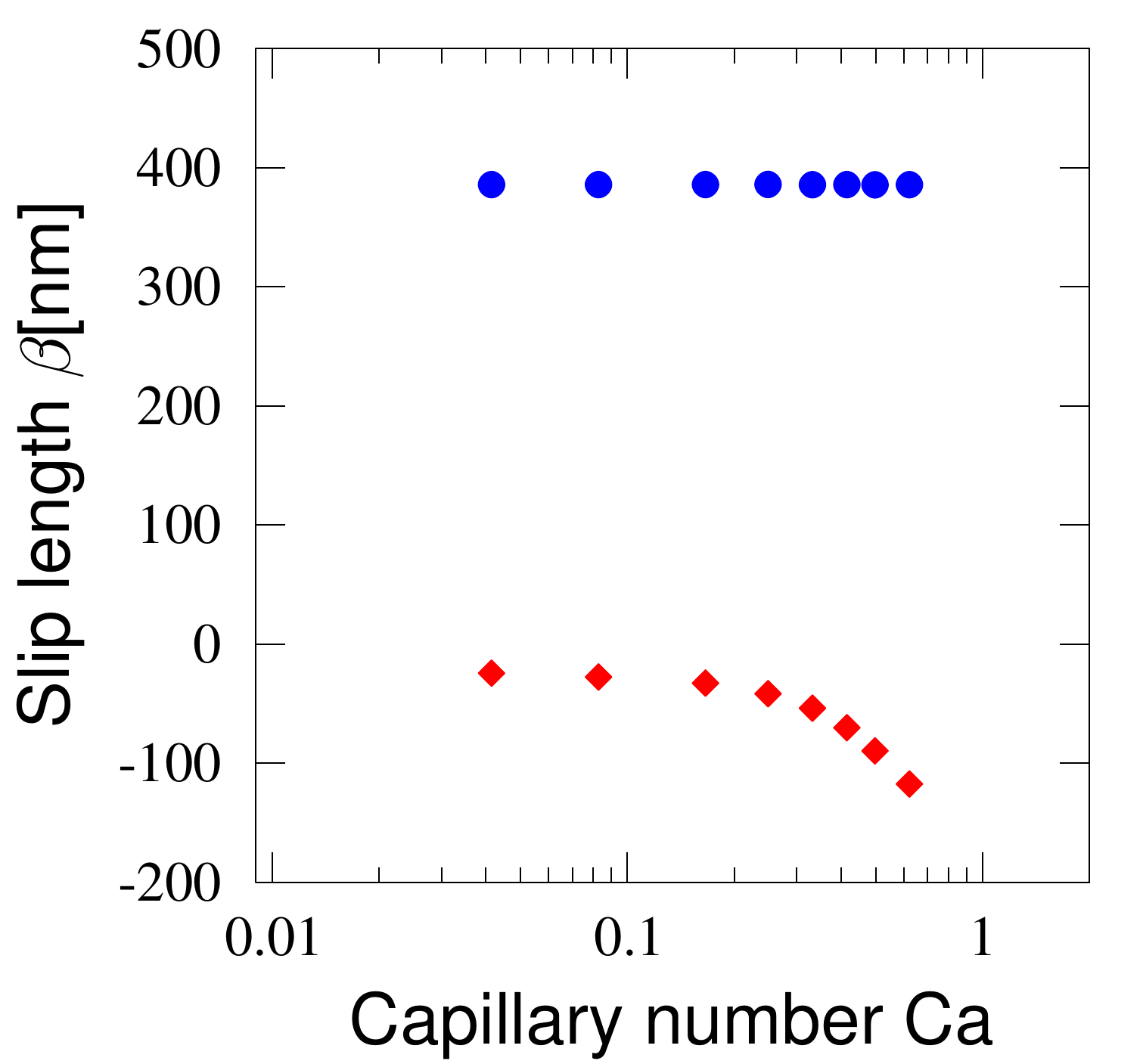}}
\caption{\label{fig:vis}
The left figure shows a snapshot of a bubble deformed by shear flow. In
the centre, the slip length as a function of the capillary
number for a square array of bubbles with three different
protrusion angles, $\varphi = 63^\circ, 68^\circ,$ and
$71^\circ$ (from uppermost to lowermost) is shown.
The inset shows cross sections of liquid-gas interfaces
for four capillary numbers~\cite{bib:jens-jari:2008}.
The right figure shows the slip length as a function of
capillary number for a surface with cylindrical bubbles.
Circles denotes the values for flow parallel to the bubbles
and diamonds for the perpendicular direction.
}
\end{figure*}

Next, the shear-rate dependence of the slip length is investigated.  As
the shear rate and thus the viscous stresses grow the bubbles are deformed
(see Fig.~\ref{fig:vis}, left) and the flow field is modified.  In the
central part of Fig.~\ref{fig:vis}, we show the simulated slip length as a
function of the Capillary number for three different protrusion angles. 
The Capillary numbers chosen are in higher end of the experimentally available
range. Our results show shear-rate dependent slip, but the behavior is
opposite to that found in some experiments: in fact, the slip lengths
measured by us decrease with increasing shear due to a deformation of the
bubbles.  In the experiments, surface force apparatuses are used (see,
e.g., Ref.~\cite{bib:zhu-granick-01}), where a strong increase in the slip
is observed after some critical shear rate. This shear-rate dependence has
been explained, e.g., with formation and growth of
bubbles~\cite{bib:degennes-02,lauga-brenner-04}. In our simulations,
there is no formation or growth of the bubbles as we only simulate a
steady case for given bubbles. The experiments on the other hand are
dynamic. However, our results indicate that the changes in the flow field
which occur due to the deformation of the bubbles cannot be an explanation
for the shear-rate dependence observed in some experiments. Our results
are consistent with~\cite{bib:jens-kunert:2007b} and the previous section,
where it is shown that smaller roughness leads to smaller values of a
detected slip. In the present case, the shear reduces the average height
of the bubbles and thus the average scale of the roughness decreases as
well.
 
Finally, we consider a surface patterned with grooves. Cylindrical bubbles
protrude to the flow channel from these holes with protrusion angle $\varphi =
72^\circ$, and the area fraction of slots is 0.53. We apply shear both parallel
and perpendicular to the slots. The slip length is strongly dependent on the
flow direction~\cite{bib:jens-jari:2008}. For parallel flow the slip length is
positive, but for the perpendicular case it becomes negative. Flow direction
affects also greatly on the shear-rate dependence (cf. Fig.~\ref{fig:vis},
right). When flow is parallel to the grooves no shear-rate dependence is
observed, but for the perpendicular case this dependence is similar to that
seen on hole arrays.  These results can be understood on the basis of deforming
bubbles. For perpendicular flow the bubbles are able to deform, but for the
parallel case the bubbles retain their shape regardless of the shear rate.

\section{Conclusion}
\REV{ED}{In this paper we review applications of the lattice Boltzmann method
to microfluidic problems. The main focus of the paper is on our own research related to the
validation of the no-slip boundary condition. By introducing a model
for hydrophobic fluid-surface interactions and studying pressure-driven
flow in microchannels, we show that an experimentally detected slip can
have its origin in hydrophobic interactions, but is constant with varied
shear rates and decreases with increasing pressure. Another effect that
was not fully understood so far is the influence of surfaces roughness. We
are able to apply our simulations to surface data obtained from AFM
measurements of experimental samples. We show that ignoring roughness can
lead to large errors in a detected slip. In fact, we propose that
roughness alone could often be the reason for apparent boundary slip. 

Microscale bubbles at surfaces allow to tailor the slip properties of a
surface. Such a surface with bubbles may yield negative slip, i.e., increased
resistance to flow, if bubbles are strongly protruding to the channel. The
lattice Boltzmann simulations capture the deformability of bubbles and thus
allow to study the influence of the shear rate on the deformation of the
interface and it's effect on the measured slip. We find that the slip decreases
with increasing shear rate demonstrating that shear induced bubble deformation
cannot explain recent experimental findings where slip increases with
increasing shear rate.

In the current review, we also demonstrate the suitability of the lattice
Boltzmann method for modeling microfluidic applications: in contrast to
molecular dynamics, it is able to reach experimentally available time and
length scales. This allows one to compare simulation results to experimental
data directly as demonstrated in the case of simulations of flow along surface
data obtained from AFM measurements of ``real'' samples.} 

\section*{ACKNOWLEDGEMENTS}
This work was financed within the DFG priority program ``nano- and
microfluidics'', the collaborative research center 716, the German
academic exchange service (DAAD), and by the
``Landesstiftung Baden-W\"urt\-tem\-berg''.  We thank the Neumann Institute
for Computing, J\"{u}lich and the Scientific Supercomputing Center,
Karlsruhe for providing the computing time and technical support for the
presented work. Jyrki Hokkanen (CSC -- Scientific Computing Ltd., Espoo,
Finland) is gratefully acknowledged for the bubble visualizations. 
Christian Kunert acknowledges fruitful discussions with P.~Szymczak.


\end{document}